\documentclass[runningheads,a4paper]{llncs}

\usepackage{amssymb}
\usepackage{graphicx}

\usepackage{algorithm}
\usepackage[noend]{algorithmic}

\usepackage{url}
\usepackage{hyperref}

\urldef{\mailsa}\path|{clifton.phua, yuzhang.feng, junyao.ji, timothy.soh}@sas.com|
\newcommand{\keywords}[1]{\par\addvspace\baselineskip
\noindent\keywordname\enspace\ignorespaces#1}

\begin{document}

\mainmatter  % start of an individual contribution

\title{Visual and Predictive Analytics \\on Singapore News: \\Experiments on GDELT, Wikipedia, and \^{}STI}

\titlerunning{Visual and Predictive Analytics on Singapore News}

\author{Clifton Phua \and Yuzhang Feng\and Junyao Ji\and Timothy Soh}
\authorrunning{Clifton Phua \and Yuzhang Feng\and Junyao Ji\and Timothy Soh}

\institute{SAS Institute Pte Ltd\\
20 Anson Road, Level 8\\
Singapore 079912\\
\mailsa\\
\url{http://www.sas.com/singapore}
}


\toctitle{Visual and Predictive Analytics on Singapore News: Experiments on GDELT, Wikipedia, and \^{}STI}
\tocauthor{Clifton Phua and Yuzhang Feng and Junyao Ji and Timothy Soh}
\maketitle

\begin{abstract}
The open-source Global Database of Events, Language, and Tone (GDELT) is the most comprehensive and updated Big Data source of important terms extracted from international news articles . We focus only on GDELT's Singapore events to better understand the data quality of its news articles, accuracy of its term extraction, and potential for prediction. To test news completeness and validity, we visually compared GDELT (Singapore news articles' terms from 1979 to 2013) to Wikipedia's timeline of Singaporean history. To test term extraction accuracy, we visually compared GDELT (CAMEO codes and TABARI system of extraction from Singapore news articles' text from April to December 2013) to SAS Text Miner's term and topic extraction. To perform predictive analytics, we propose a novel feature engineering method to transform row-level GDELT from articles to a user-specified temporal resolution. For example, we apply a decision tree using daily counts of feature values from GDELT to predict Singapore stock market's Straits Times Index (\^{}STI). Of practical interest from the above results is SAS Visual Analytics' ability to highlight the various impacts of June 2013 Southeast Asian haze and December 2013 Little India riot on Singapore. Although Singapore is unique as a sovereign city-state, a leading financial centre, has strong international influence, and consists of a highly multi-cultural population, the visual and predictive analytics reported here are highly applicable to another country's GDELT data.
\end{abstract}

\section{Introduction}

The Global Data on Events, Location and Tone (GDELT) provides one of the best opportunities to perform Big Data analytics on news and events \cite{arva2013improving}. GDELT \cite{leetaru2013gdelt}:
\begin{itemize}
\item is freely available and uses a variety of international news sources with daily updates
\item contains more than 250 million events from 1979 to present
\item is CAMEO-coded \cite{gerner2002conflict}, uses TABARI system for events \cite{best2013analysis},  \href{http://www.geonames.org/}{GeoNames} for geocoding, and regularly introduces new enhancements to the data
\end{itemize}

We focus our analytical efforts only on Singapore which is a relatively small and peaceful country in the world, compared to parts of the Middle East\footnote{How Computers Can Help Us Track Violent Conflicts - Including Right Now in Syria, \url{http://themonkeycage.org/2013/07/09/how-computers-can-help-us-track-violent-conflicts-including-right-now-in-syria/} (2013)} \footnote{The Arab Spring and GDELT, \url{http://blog.gdelt.org/2013/10/02/the-arab-spring-and-gdelt/}} and Africa \cite{perry2013machine}. One benefit is that, as authors of this paper, we know recent Singapore's key events well; and another benefit is that we have to experiment with machine learning (such as building of decision trees) beyond conflict, violence, or protests using GDELT. Below are 4 questions and our corresponding answers after performing initial experiments:
\begin{enumerate}
\item Can GDELT be used to understand Singapore better? Yes, to a certain extent. We were able to highlight some key events by exploring different analytical approaches
\item Is GDELT's news complete and valid? The news sources are comprehensive, but not all articles were valid for our analytical objectives
\item Can GDELT's term extraction be improved? There can be additional insights from using natural language processing, on top of CAMEO codes and TABARI system
\item Can GDELT be used to predict non-political outcomes? We propose a novel feature engineering algorithm to transform GDELT into a machine learning dataset. For experimental purposes, we attempt to naively predict an index of the Singapore stock market
\end{enumerate}

\begin{figure}
\centering
\includegraphics[width=12cm]{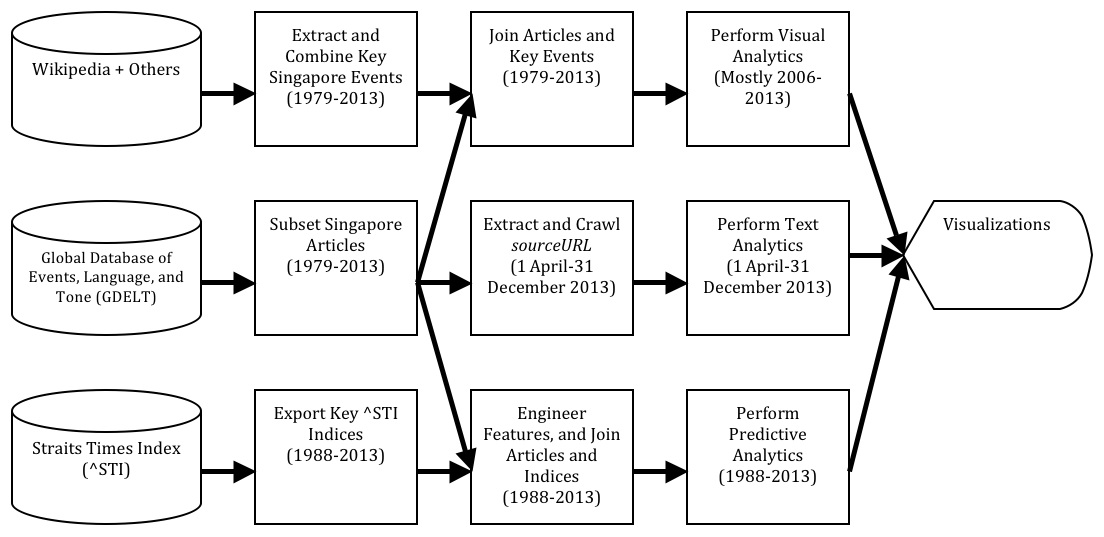}
\caption{Overview of databases, data preparation, and analytical approaches.}
\label{fig:overview}
\end{figure}

With reference to 1st and 2nd columns of Figure \ref{fig:overview}, we retrieved only Singapore GDELT articles using \textit{ActionGeo\_CountryCode}\footnote{How to both download and read GDELT using just R, \url{http://blog.gdelt.org/2013/09/02/how-to-both-download-and-read-gdelt-using-just-r/}}. Next, we extract and merge key Singapore events (ground truth) from Wikipedia's \href{http://en.wikipedia.org/wiki/Timeline_of_Singaporean_history}{Timeline of Singaporean history}, Wikipedia's \href{http://en.wikipedia.org/wiki/List_of_major_crimes_in_Singapore}{List of major crimes in Singapore}, \href{http://www.bbc.co.uk/news/world-asia-15971013}{BBC Singapore chronology of key events}, and \href{http://www.timelinesdb.com/listevents.php?subjid=130\&title=Singapore}{TimelinesDB Singapore timeline}. Third, we exported the Straits Times Index (\^{}STI) from \href{http://sg.finance.yahoo.com/q/hp?s=\%5ESTI}{Yahoo Finance}. We chose \^{}STI as target variable, over interest rates/flat resale price index/unemployment rate/GDP per capita, because it has daily data and it can be slightly influenced by the previous day's news.

At 3rd column of Figure \ref{fig:overview}, we joined articles and key events (1979-2013) into a 717,124-row dataset if they shared the same date. Next, we crawled the \textit{sourceURL} column's hyperlinks (1 April 2013 to 31 December 2013) using boilerplate detection \cite{kohlschutter2010boilerplate};  and managed to extract 30\% of URLs (8,498 out of 28,177). Third, we use $DERIVEFEATURES$ Algorithm (Sub-section \ref{subsec:feature}) to engineer features, joined them to \^{}STI indices (1988-2013) with a 1-day lag, to create a 68,514-row dataset.

The last 2 columns in Figure \ref{fig:overview} show that visual analytics (Section \ref{sec:data}) and predictive analytics (Section \ref{sec:potential}) were performed. The visualizations are possible using \href{http://www.sas.com/en_us/software/business-intelligence/visual-analytics.html}{SAS Visual Analytics} (VA) \cite{sas2013va}. VA provides a robust set of business intelligence capabilities and approachable analytics, enabling different types of users to gain insights from any size of data through data visualization and exploratory analysis. Text analytics (Section \ref{sec:accuracy}) was performed using \href{http://www.sas.com/en_us/software/analytics/text-miner.html}{SAS Text Miner} (TM) \cite{sas2012tm}. TM discovers information buried in collections of text. By automatically reading text data and delivering algorithms for rigorous, advanced analyses, the solution makes it possible to grasp future trends and act on new opportunities more precisely and with less risk.

%\section{Background}

\section{Data Quality of News Articles}
\label{sec:data}

Due to space constraints, Figures \ref{fig:monthly-articles-count} to \ref{fig:geomap-november-2013} in Appendix provides sufficient background to our choice of bubble plot and geomap analysis to understand GDELT's data quality.

\subsection{Bubble Plot Analysis}

\begin{figure}
\centering
\includegraphics[width=12cm]{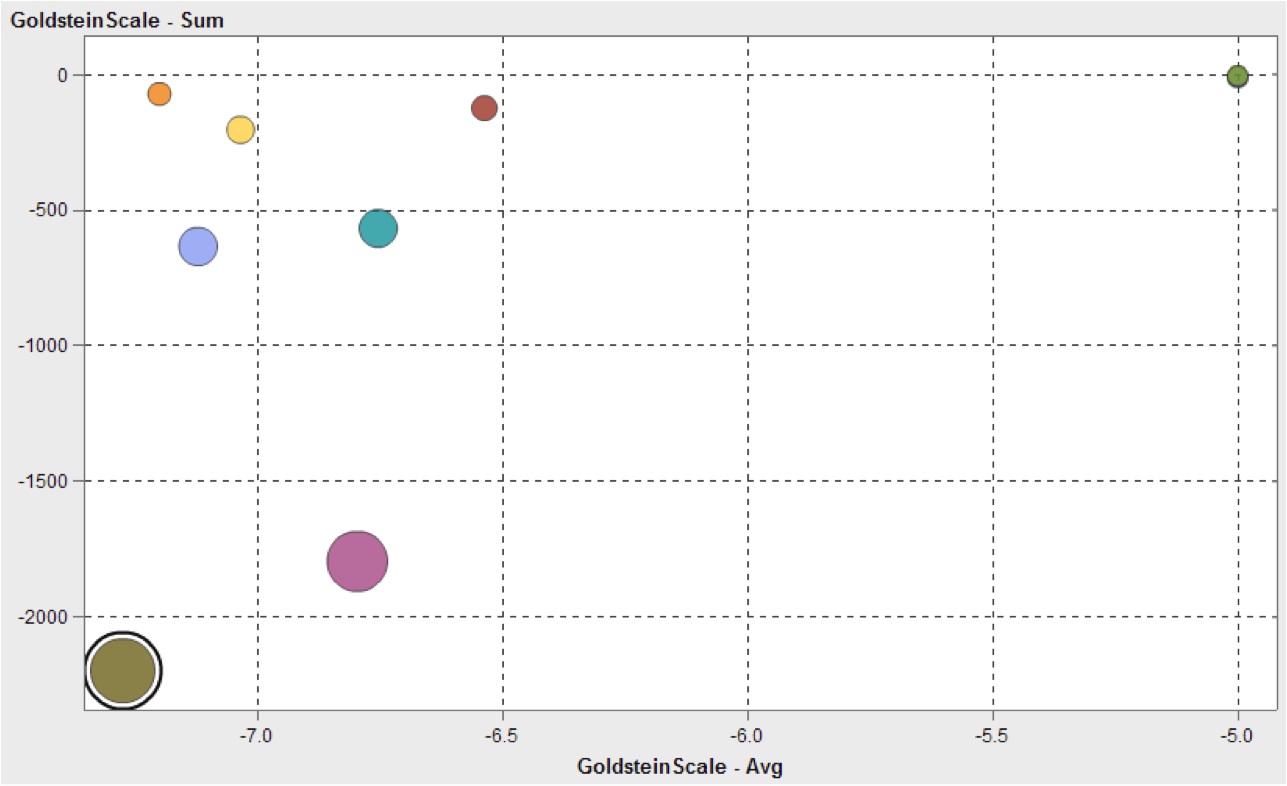}
\caption{Bubble plot on monthly Singapore GDELT articles from 1 April to 31 December 2013.}
\label{fig:gdelt-wikipedia-comparison}
\end{figure}

Figure \ref{fig:gdelt-wikipedia-comparison} shows a bubble plot from April to December 2013 (\textit{sourceURL} only existed from April 2013) with the following criteria:
\begin{enumerate}
\item \textit{QuadClass = Material Conflict}
\item \textit{event = 1} (months where key Singapore events occurred)
\item \textit{IsRootEvent = true} (GDELT events were parsed in the first paragraph of the news article)
\item Only regional news publications from Singapore/Malaysia/Indonesia were chosen (\url{www.channelnewsasia.com}, \url{news.asiaone.com}, \url{www.straitstimes.com}, \url{sg.news.yahoo.com}, \url{www.todayonline.com}, \url{www.businesstimes.com.sg}, \url{www.singaporestar.com}, \url{www.thestar.com.my}, \url{www.themalaysianinsider.com}, \url{www.themalaymailonline.com}, \url{www.nst.com.my}, and \url{www.thejakartapost.com} - that were parsed from the \textit{sourceURL} column, to eliminate over-reporting of events that are of international interest which would overshadow other key Singapore events). The selection of news sources from \textit{sourceURL} column is only for the bubble plot. The other visualizations used all news sources, because the amount of data was already small from applying other filters.
\end{enumerate}

%This bubble plot in Figure \ref{fig:gdelt-wikipedia-comparison} is based on a combined dataset of Singapore GDELT articles with key Singapore events as ground truth, where there is an additional column named \textit{event} with value of 1 if Singapore GDELT article occurs within a time window of the key Singapore event (7 days before or 14 days after).
Big bubbles at bottom left or top right ends of the plot are interesting. For example, the highlighted bubble at the bottom left is Singapore news articles in July 2013, upon closer inspection of its \textit{sourceURL} column, show a large number of news sources linking to the locally well-known Kovan double murder on 12 July 2013.

Visually, we see only a small number of the key Singapore events clearly highlighted in the GDELT \textit{GoldsteinScale} trends (large bubbles on the extreme ends). This is usually caused by many other less important and random events occurring near or during the duration (7 days before or 14 days after) of the key Singapore events. To test the validity of Singapore GDELT data, a much more detailed list of key Singapore events is required. %The other small bubbles would be overshadowed by other bubbles with the year filter off.

%\begin{figure}
%\centering
%\includegraphics[width=12cm]{GDELT-GoldsteinScale-vs-Wikipedia-GroundTruth-Bubble-Plot-Comparison-sourceURL.jpg}
%\caption{\textbf{(SHOULDN'T THIS BE A TABLE???)}}
%\label{fig:xxx1}
%\end{figure}

\subsection{Geomap Analysis}

\begin{figure}
\centering
\includegraphics[width=12cm]{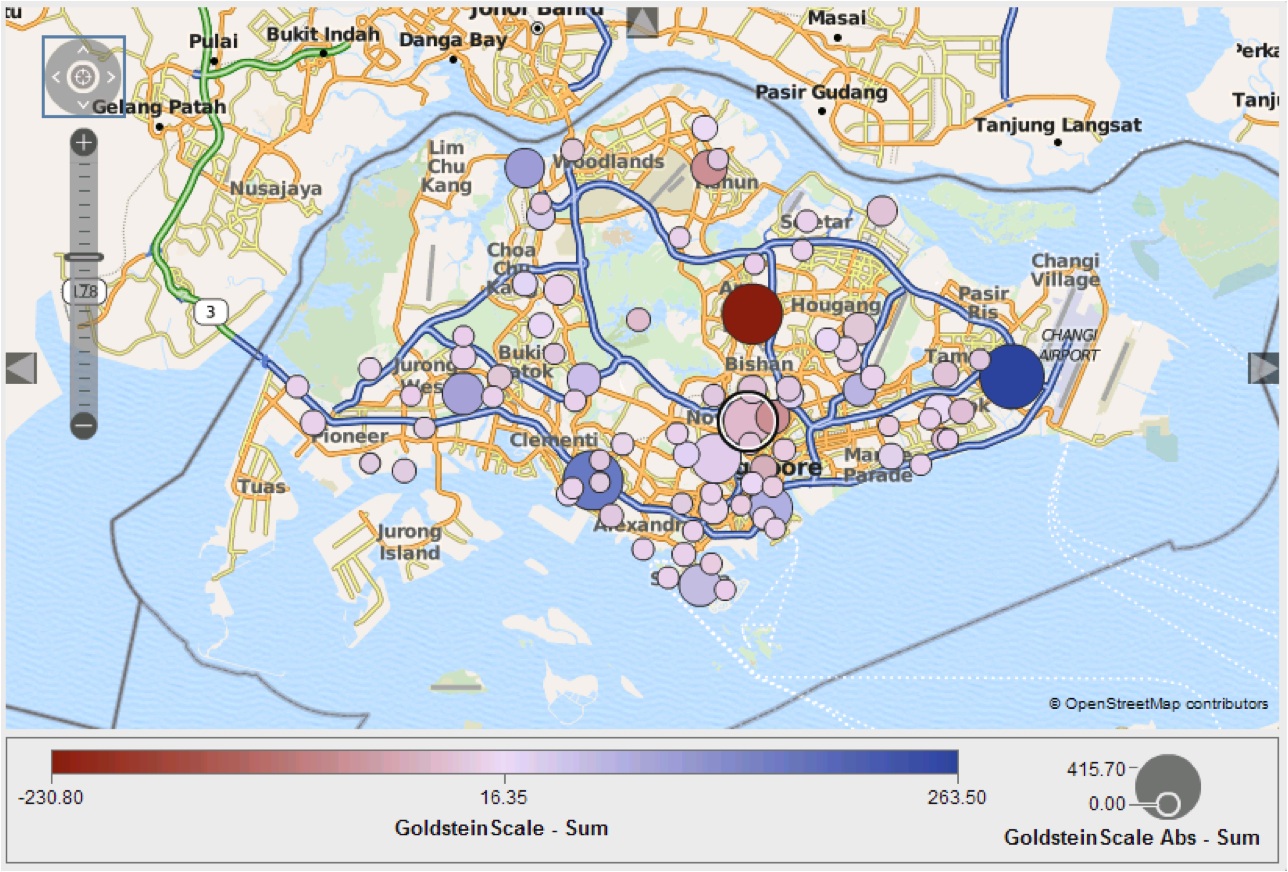}
\caption{Geomap of Singapore GDELT articles from December 2013 (1 month only).}
\label{fig:geomap-december-2013}
\end{figure}

Figure \ref{fig:geomap-december-2013} is the geomap of Singapore GDELT articles from December 2013 (1 month only). The highlighted circle corresponds to GDELT articles about the Little India riot which happened on 8 December 2013 - it was the first Singapore riot in over 40 years. This circle does not stand out very much from among others. This shows that not all significant events in Singapore can be clearly discovered using GDELT.

%normalize [number of protests] by the total number of events per country over the same time periods\footnote{Normalizing GDELT Protest Data, \url{http://blog.gdelt.org/2013/09/28/normalizing-gdelt-protest-data/}}

\section{Accuracy of Term Extraction}
\label{sec:accuracy}

Although TABARI provides effective machine coding using lead sentences, one of its weakness is the dependence on details available in the actor dictionaries \cite{best2013analysis}. We propose the following use of natural language processing to complement machine coding.

From \textit{sourceURL} column in our Singapore GDELT dataset, we are able to retrieve a set of news articles' text from April 2013 to December 2013. We perform two text analytics tasks on this set of articles, namely concept link exploration
and topic analysis.

Due to space constraints, Tables \ref{table:topic25} to \ref{table:common-topic} are available in the Appendix.

\subsection{Concept Link Exploration}

Using the concept link exploration feature, we are
able to visually analyze which terms co-occur with a given term
most frequently in the same document. The terms are visualized
using nodes and the thickness of the links represent the degree
of co-occurrence.

% DO NOT UNCOMMENT THIS BLOCK
%To add in ta-shanmugam.jpg
%\begin{figure}
%\centering
%\includegraphics[width=1.0\columnwidth]{Accuracy-of-term extraction/ta-shanmugam.jpg}
%\caption{Concept link illustrating term association for the term Shanmugam}
%\label{fig:ta-shanmugam}
%\end{figure}

% DO NOT UNCOMMENT THIS BLOCK
%To add in ta-government.jpg
%\begin{figure}
%\centering
%\includegraphics[width=1.0\columnwidth]{Accuracy-of-term extraction/ta-government.jpg}
%\caption{Term association for government}
%\label{fig:ta-government}
%\end{figure}

%To add in ta-oil.jpg
\begin{figure}
\centering
\includegraphics[width=12cm]{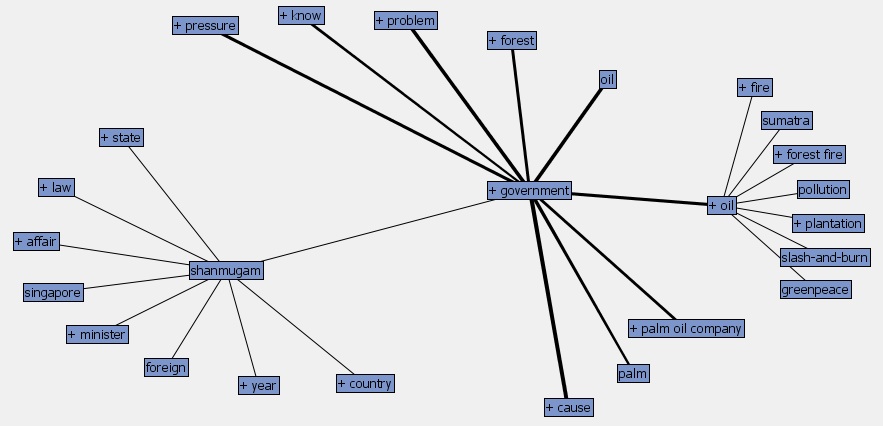}
\caption{Concept link illustrating term association for ``Shanmugam", ``government", and ``oil".}
\label{fig:ta-oil}
\end{figure}

The term ``Shanmugam'' is one of the most frequent terms in our set of articles. A
politician and lawyer, K. Shanmugam is Singapore's Minister for Foreign Affairs
and Minister for Law who frequently gives public speeches and interviews. We study the term ``Shanmugam'' in more detail using concept link exploration.

Figure \ref{fig:ta-oil} illustrates the process and the output. We start
by visualizing the top 9 terms that are most closely related to the term ``Shanmugam''.
Among these terms, we are most interested in the ``government'' and explore further
by expanding it in the concept link graph. It is easy to identify that many terms
that have strong co-occurrence with the term ``government'' are closely related to
``oil''. If we explore the term ``oil'' further, we find terms
that are related to Indonesian forest fires which resulted in the Southeast Asian haze - it caused record high levels of air pollution in Singapore of which some days were hazardous to health. Putting
them together, it makes sense because Shanmugam was the main politician who gave
public statements to the Indonesian government, and requested cooperation and collaboration.

\subsection{Topic Analysis}

Topic analysis provides ways to define, discover and modify sets of topics contained
in the collection of texts. A topic is defined by a set of terms that are strongly
associated with a subset of the text collection. Each document can contain zero, one,
or many topics. Terms that are used to describe and define one topic can be used in
other topics.

To conduct topic analysis, the number of topics has to be provided as an input parameter. Heuristically, we choose to
generate 25, 50, 75 and 100 topics for each run. The results of 25-topic analysis
is shown in Appendix's Table \ref{table:topic25}. Due to space constraints, we show, among
many others, 3 key variables for each topic including the top 5 topic terms, the
number of topic terms, and the number of documents in the topic. We provided a few key Singapore events in the ``Description" column.

We also conducted similar studies for 50, 75 and 100 topics. Due to space constraints, we
omit the detailed results. Across the four sets of results, we have other interesting
observations. We are able to identify 8 breaking news from the set of articles as
shown in Appendix's Table \ref{table:breaking}. The topics on Little India riot and Southeast Asian haze exist in all the four result sets. In addition, we discover more
topics common to all four sets of results as shown in Appendix's Table \ref{table:common-topic}.
Even with much increase in the number of topics to be identified, the
number of common topics does not increase as much.

We have also performed topic analysis on Singapore GDELT articles in quarters 3 and 4 of 2013
separately. Our results show that the haze topic has gone down from 11th place in
quarter 3 to 14th in quarter 4 in terms of number of documents on the topic. Over the
same period, it can be observed that the topic of Little India riot emerged quickly
to 11th in quarter 4. These results are consistent with the chronological order of the
two key Singapore events (Southeast Asian haze in June and Little India riot in December 2013).

\section{Potential for Prediction}
\label{sec:potential}

\newcommand{\derive}{\ensuremath{\mbox{\sc DeriveFeatures}}}

\subsection{Feature Engineering using $\derive$ Algorithm}
\label{subsec:feature}

In most transactional datasets, there might be multiple records which correspond
to a single temporal unit.
%For example, in GDELT, there are typically many
%event records for a given day.
%In order to create the mining view as a daily
%view, the first requirement is to have one, and only one, record per day.
To roll up the multiple records on the daily level to have exactly one record
per day, we need to create a set of
new variables to represent the combination of the original variable and its
corresponding values. For example, when working with the GDELT dataset, it is
often useful to aggregate the records and find out how many events with the
value \textit{Material Conflict} for the \textit{Quadclass} variable happened on a given
day. Such derived information can be used as an important feature for various
data analytics/mining tasks such as predictive modeling and clustering. Unfortunately,
it is often tedious and error-prone to generate such features manually. Here, we describe a novel and generic feature engineering method.

\begin{algorithm}[t!]
  \caption{$\derive(s)$}
  \begin{algorithmic}[1]
    \REQUIRE A dataset $s$ with variables $V_0, \ldots, V_n$ of which $V_0$
    is a date/time variable which has $m$ distinct values, and $V_1, \ldots, V_n$
    are either nominal or numeric variables
    \ENSURE Returns a dataset $s'$ of $m$ observations, one for each distinct
    value of $V_0$, and $n$ sets of variables $F_1, \ldots, F_n$
    %one set for each
%    of $V_1, \ldots, V_n$ such that each set contains a set of feature variables
%    derived from their corresponding original variable
%    \medskip
    \STATE $s' \leftarrow$ a dataset containing only the $m$ distinct observations of variable $V_0$
    \FORALL {$i = 1$ to $n$ }
        \IF {$V_i$ is a nominal variable}
            \STATE $d_i \leftarrow$ distinctValues($V_i$)
            \STATE Expand $s'$ by adding $d_i + 1$ variables $V_i\_0, \ldots, V_i\_{d_j}$
            \FORALL {$j = 0$ to $d_i$ }
                \FORALL {$v_0 \in V_0$ }
                    \IF {$j = 0$}
                        \STATE $V_i\_j(v_0) \leftarrow$ selectCount($v_0$, $V_i$, MissingValue)
                    \ELSE
                        \STATE $V_i\_j(v_0) \leftarrow$ selectCount($v_0$, $V_i$, ${v_i}_j$)
                    \ENDIF
                \ENDFOR
            \ENDFOR
        \ELSIF {$V_i$ is a numeric variable}
            \STATE $(b_1, \ldots, b_l) \leftarrow$ binning($V_i, l$)
            \STATE Expand $s'$ by adding $d_i + 1$ variables $V_i\_0, \ldots, V_i\_l$
            \FORALL {$k = 0$ to $l$ }
                \FORALL {$v_0 \in V_0$ }
                    \IF {$k = 0$}
                        \STATE $V_i\_k(v_0) \leftarrow$ selectCount($v_0$, $V_i$, MissingValue)
                    \ELSE
                        \STATE $V_i\_j(v_0) \leftarrow$ selectCountFromBin($v_0$, $V_i$, $b_k$)
                    \ENDIF
                \ENDFOR
            \ENDFOR
        \ENDIF
    \ENDFOR
    \RETURN $s'$
  \end{algorithmic}
  \label{alg:derive}
\end{algorithm}

Algorithm \ref{alg:derive} assumes
that the input dataset $s$ contains $n+1$ variables $V_0, \ldots, V_n$ of which $V_0$ is
a date/time variable which has $m$ distinct values, and $V_1, \ldots, V_n$ are either
nominal or numeric variables. The output dataset $s'$ contains $m$ observations, one for
each distinct value of $V_0$. The variables of $s'$ include $V_0$ and another $n$ sets
of variables $F_1, \ldots, F_n$, one set for each of $V_1, \ldots, V_n$ such that each
set contains a set of feature variables derived from their corresponding original variable.

The output dataset $s'$ is constructed as follows. For each nominal variable $V_i$ (lines 3
to 11), we find out the number of distinct values ($d_i$) for the variable in the input
dataset (line 4) and create $d_i + 1$ new variables for the output dataset (line 5). Then
the value of the variable for the time/date unit in the output dataset is the number of
records with that particular value for the variable for the time/date unit in the input
dataset (line 11). We need one extra variable to keep track of the number of records with
a missing value for the variable for the time/date unit in the input dataset (line 9).
%For
%practicality it is sometimes sufficient to look at the top $\alpha$ most frequent values
%of the variable rather than all the distinct values of the variable.
We perform the similar
procedures for numeric variables (lines 12 to 20), except that for each of the variables we
create $l$ bins by using either equal-interval binning or equal-frequency binning method on
the all the values of the variable. Similarly we create $l+1$ new variables for the output
dataset, one for each bin and the extra one for those records with a missing value. For
consistency, we have illustrated the algorithm using the same selection function as the one
we used for nominal variables: more specifically, counting the number of records in the
corresponding bin for the given date/time unit in the input file. To create other possibly
useful features, summation and averaging can also be used as alternative ways for aggregation.

\subsection{Prediction using Decision Tree Algorithm}

Due to space constraints, Figure \ref{fig:overall-decision-tree} is available in the Appendix.

\begin{figure}[htb]
\centering
\begin{minipage}[b]{0.45\linewidth}
\includegraphics[width=4.5cm]{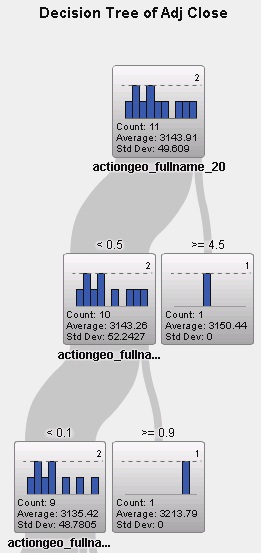}
\caption{Decision tree constructed using independent variables of Singapore GDELT articles at day $t$ and target variable of STI adjusted close price at day $t+1$ from 13 June 2013 to 27 June 2013 (2 weeks). The Southeast Asian haze started from 13 June 2013 and subsided around 27 June 2013.}
\label{fig:haze-decision-tree}
%\begin{figure}
%\centering
%\includegraphics[width=12cm]{NodeDetails2_HAZE.jpg}
%\caption{\textbf{(SHOULDN'T THIS BE A TABLE???)}}
%\label{fig:xxx1}
%\end{figure}
\end{minipage}
\quad
\begin{minipage}[b]{0.45\linewidth}
\includegraphics[width=4.5cm]{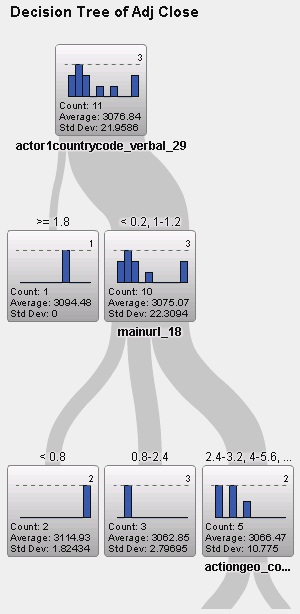}
\caption{Decision tree constructed using independent variables of Singapore GDELT articles at day $t$ and target variable of STI adjusted close price at day $t+1$ from 8 December 2013 to 22 December 2013 (2 weeks). The Little India riot happened on 8 December 2013 and the media coverage continued approximately for the next 2 weeks.}
\label{fig:riot-decision-tree}
%\begin{figure}
%\centering
%\includegraphics[width=12cm]{NodeDetails2_RIOT.jpg}
%\caption{\textbf{(SHOULDN'T THIS BE A TABLE???)}}
%\label{fig:xxx1}
%\end{figure}
\end{minipage}
\end{figure}

Figure \ref{fig:haze-decision-tree} shows the top-3 variables (out of 1,100), during the haze period, to predict \^{}STI adjusted close price are \textit{ActionGeo\_Fullname = Toa Payoh},  \textit{ActionGeo\_Fullname = Singapore}, and \textit{ActionGeo\_Fullname = Geylang}. Toa Payoh and Geylang are in central parts of Singapore. The former had good news coverage on a facial mask distribution drive for its aged population, and the latter had a number of reports of haze over its skyline and negative impact on outdoor businesses.

Figure \ref{fig:riot-decision-tree} shows the top-3 variables (out of 1,100), during the riot period, to predict \^{}STI adjusted close price are \textit{Actor1CountryCode = Pakistan},  \textit{mainURL = zeenews.india.com}, and \textit{ActionGeo\_CountryCode = Singapore}. It shows there was extensive news coverage of the Little India riots by Pakistani, Indian, and Singaporean media.

%\section{Discussion}

\section{Conclusion}

From this initial work on Singapore GDELT articles, we now have a better understanding of the GDELT's data quality, content, and potential applications. We will track GDELT's developments through its blog\footnote{Blogging about GDELT (gdelt.utdallas.edu), \url{http://blog.gdelt.org/}}, and future work could include forecasting using GDELT \cite{brandt2013forecasting}.

\section*{Appendix}

%\section{Appendix: Data Quality of News Articles}

\begin{figure}[htb]
\centering
\includegraphics[width=12cm]{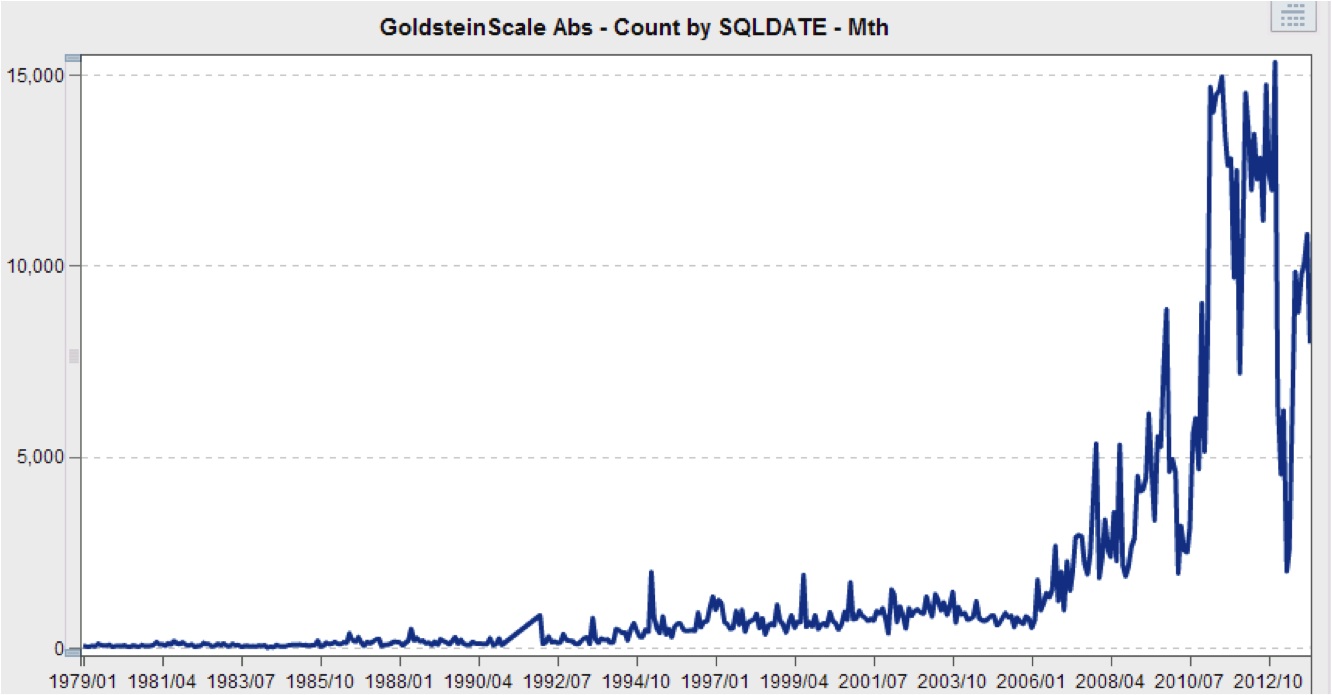}
\caption{Line chart showing number of Singapore GDELT articles from 1979-2013, aggregated by month \cite{yonamine2011working}.  It makes sense to build visualizations and analyses only from 2006-2013 because there is a significantly larger number of GDELT articles compared to 1979-2005. There are some months within 2006-2013 with much fewer Singapore GDELT articles.}
\label{fig:monthly-articles-count}
\end{figure}

\begin{figure}[htb]
\centering
\begin{minipage}[b]{0.45\linewidth}
\includegraphics[width=6cm]{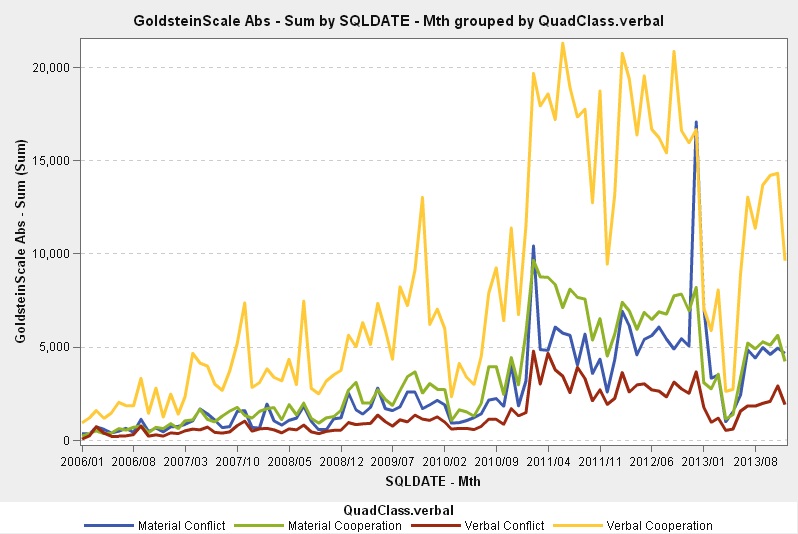}
\end{minipage}
\quad
\begin{minipage}[b]{0.45\linewidth}
\includegraphics[width=6cm]{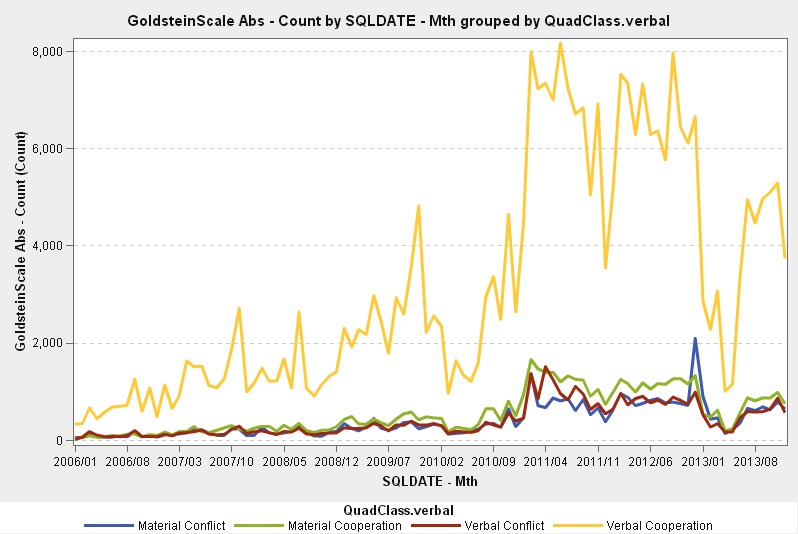}
\end{minipage}
\quad
\begin{minipage}[b]{0.45\linewidth}
\includegraphics[width=6cm]{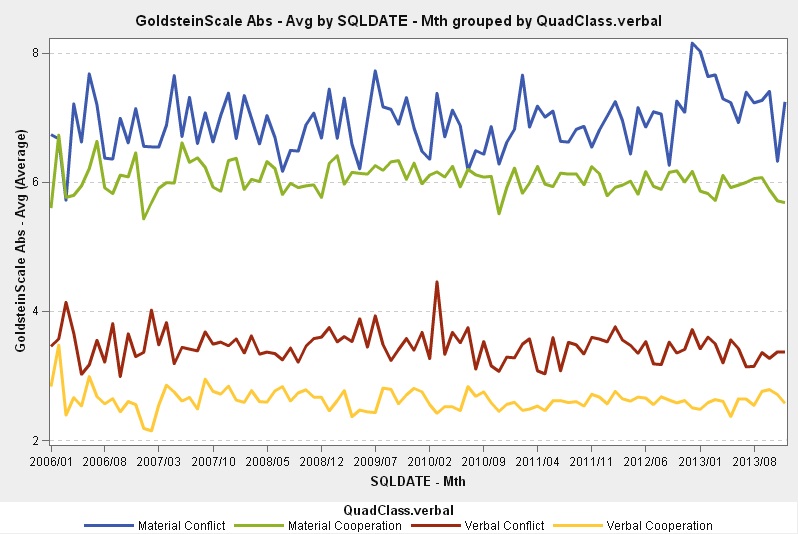}
%GoldsteinScale scale aggregations by month graphs
%The goldsteinScale values were all changed to positive values, so as to reduce the amount of space taken up on the graph by removing the negative portion of the y-axis
%Actually the goldsteinScale avg by month graph was correct, the top two lines were material cooperation and conflict, which naturally have higher values then verbal cooperation and conflict. The color scheme was a little off, because the VA exploration didnÕt allow for custom colors, so I replicated the screenshots with VA reports
\end{minipage}
\caption{Line charts, each with different aggregate functions on monthly \textit{GoldsteinScale} \cite{goldstein1992conflict}, grouped by \textit{QuadClass} \cite{leetaru2013gdelt} from 2006-2013. Top left - \texttt{Sum(GoldsteinScale)}, top right - \texttt{Count(GoldsteinScale)}, and bottom - \texttt{Avg(GoldsteinScale)}. Each of these aggregations have their own advantages and disadvantages. \texttt{Sum(GoldsteinScale)} gives the ability to observe overall impact for an aggregated time period; however, a time period with several small low-impact events can overshadow a time period with only one big event. \texttt{Count(GoldsteinScale)} shows the number of events but masks their intensity. \texttt{Avg(GoldsteinScale)} allows us to look at the impact for an aggregated time period if the variance in impact between individual events is not large; but it masks big events in a time period if there are other smaller events in the same period. Visual comparisons of all 3 line charts is necessary to evaluate the \textit{GoldsteinScale} over time. For example, according to GDELT, Singapore is a relatively peaceful country which experiences mostly \textit{Verbal Cooperation} (such as occurrence of dialogue-based meetings) and \textit{Material Cooperation} (such as receiving or sending aid). There is an obvious false positive where December 2012 \texttt{Sum(GoldsteinScale)} spike in \textit{Material Conflict} was the result of international media focus on a crime victim who was flown into Singapore's Mount Elizabeth Hospital for treatment.}
\label{fig:different-aggregates-goldstein-scale}
\end{figure}
%the disadvantages above are further exacerbated with the presence of positive and negative values. Grouping the goldstein values by quadclass can mitigate this issue, but it does not solve the problem of having to visually compare 3 different line graphs

%\begin{itemize}
%\item Verbal Cooperation: The occurrence of dialogue-based meetings (e.g. negotiations, peace talks), statements that express a desire to cooperate or appeal for assistance (other than material aid) from other actors.
%\item Material Cooperation: Physical acts of collaboration or assistance, including receiving or sending aid, reducing bans and sentencing, etc.
%\item Verbal Conflict: A spoken criticism, threat, or accusation, often related to past or future potential acts of material conflict.
%\item Material Conflict: Physical acts of a conflictual nature, including armed attacks, destruction of property, assassination, etc.
%\end{itemize}

\begin{figure}[htb]
\centering
\includegraphics[width=12cm]{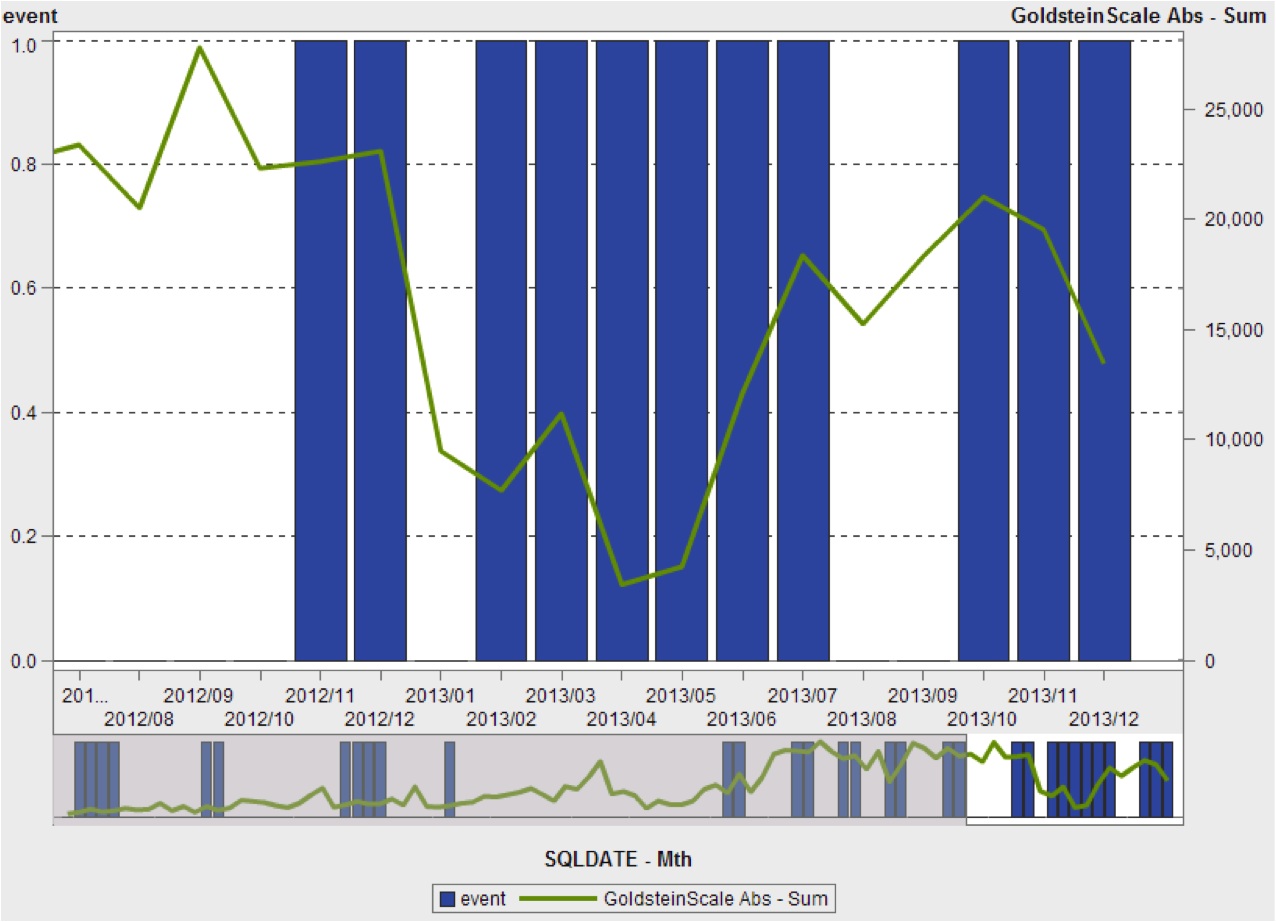}
\caption{Dual-axis bar line chart where the bar is an occurrence of a key Singapore event (such as it being listed in Wikipedia) with a 1 month window and line is the \texttt{Sum(GoldsteinScale)} on Singapore articles from 2006-2013, with focus on July 2012 to December 2013. This approach can be tedious as it requires manual trend observation; furthermore if the line is separated into 4 lines using \textit{QuadClass}, it could be more challenging to detect actual trends. Here, no obvious relationship seems to exist between ground truth events and the \textit{GoldsteinScale}.}
\label{fig:gdelt-wikipedia-window-comparison}
\end{figure}
% \textbf{(WHY NOT DAYS???)} - due to Yonamine's recommendation to use monthly data, too much data to display, and some events have not precise dates in Wikipedia
% \textbf{WHY NOT AVGTONE???}} - due to previous studies

\begin{figure}[htb]
\centering
\includegraphics[width=12cm]{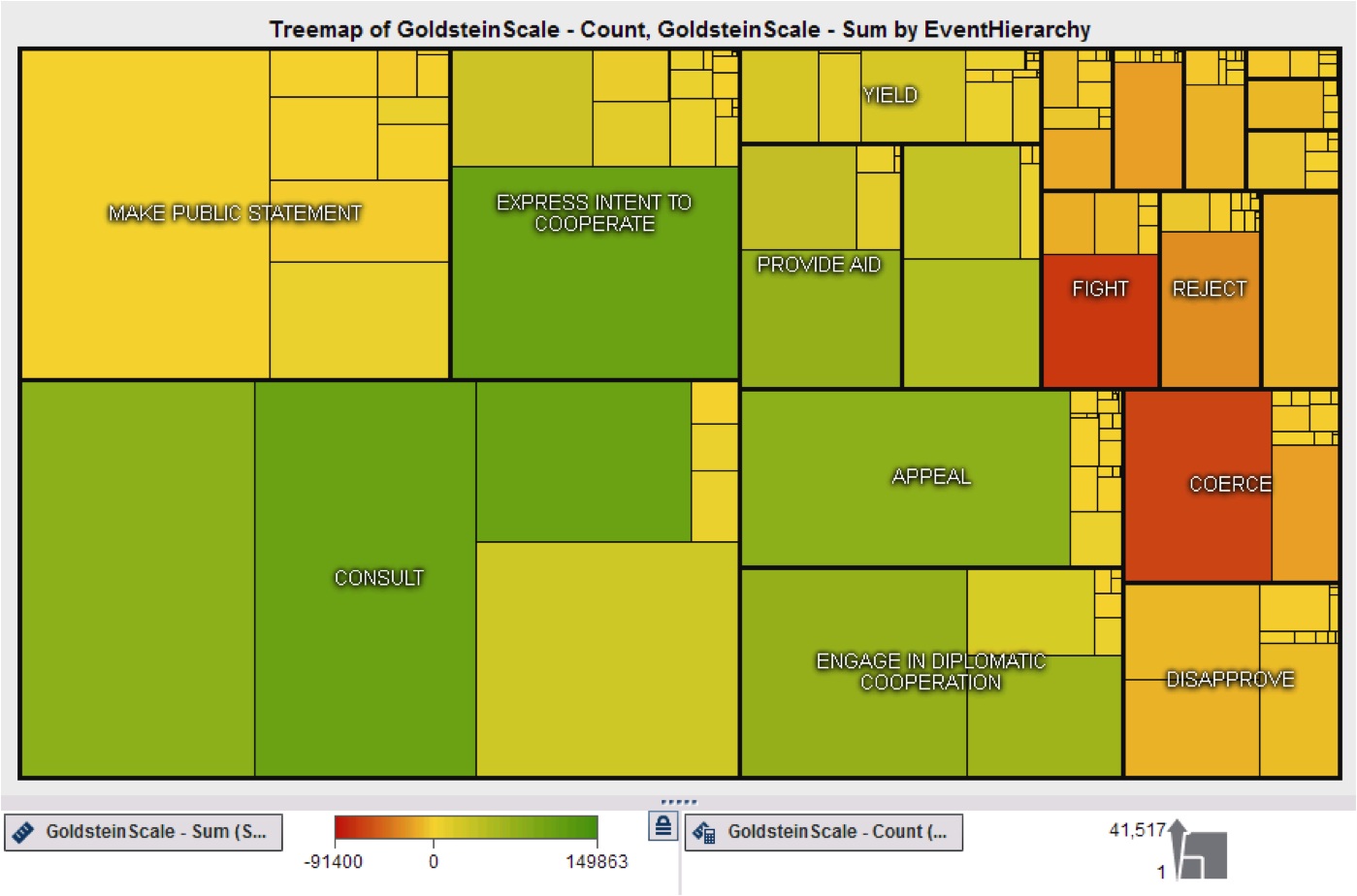}
\caption{Treemap of \textit{EventHierarchy} and \textit{GoldsteinScale} on Singapore GDELT articles from 2006-2013. \textit{EventHierarchy} consists of EventRootCode (21 distinct values), EventBaseCode (139 distinct values), and EventCode (224 distinct values). Size of each box - \texttt{Count(GoldsteinScale)} and color of each box - \texttt{Sum(GoldsteinScale)}. This treemap confirms that news about Singapore are mostly positive.}
\label{fig:treemap}
\end{figure}
%The highest level is actually QuadClass.verbal (4 distinct types), but it is not included in the hierarchy, because the treemap can only display 2 levels of the hierarchy visually and displaying 21 over 4 distinct types will make treemap more informative and visually appealing.

\begin{figure}[htb]
\centering
\includegraphics[width=12cm]{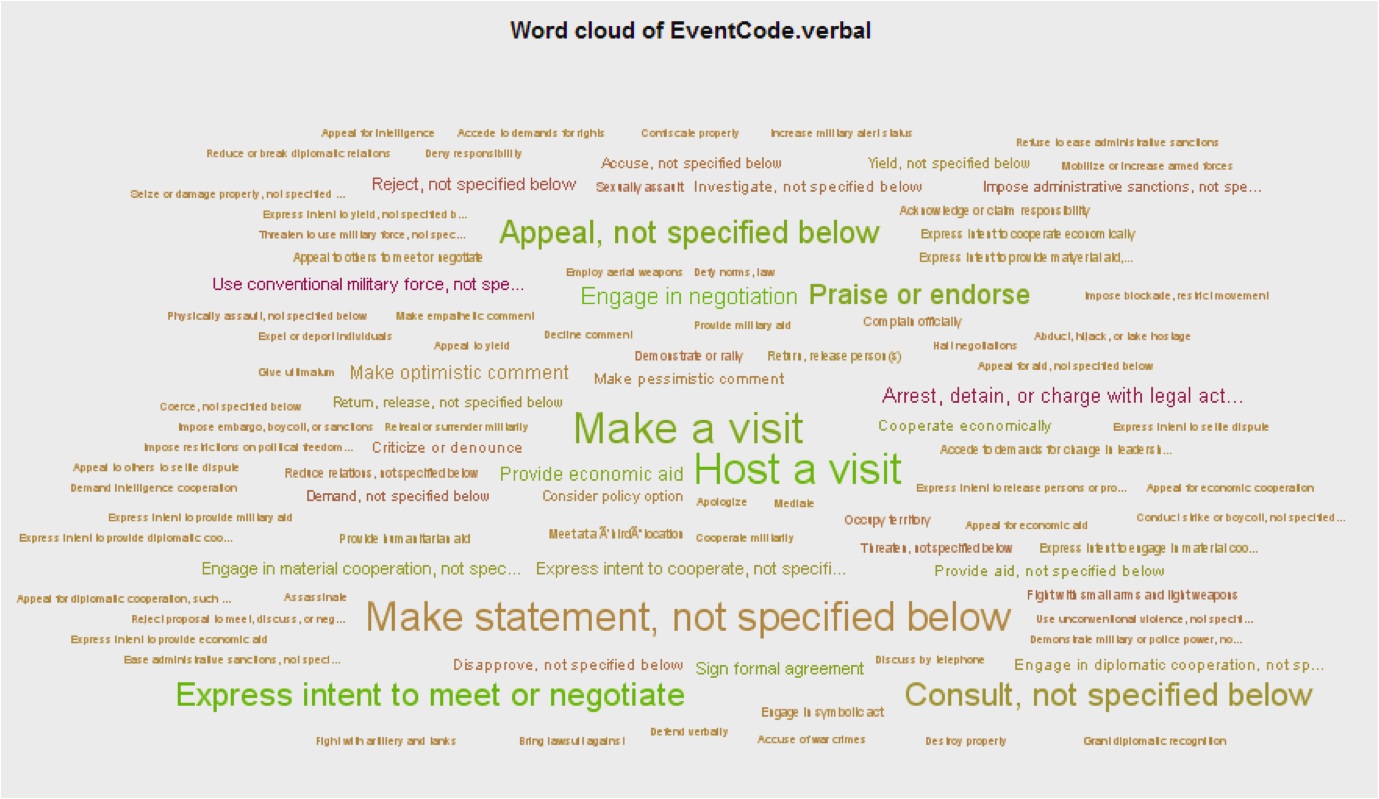}
\caption{Word cloud of \textit{EventCode} on Singapore GDELT articles from 2006-2013. Size of words - \texttt{Count(\textbf{EventCode})} and color of words - \texttt{Sum(GoldsteinScale)}. Again, this word cloud confirms that news about Singapore are mostly positive.}
\label{fig:word-cloud}
\end{figure}

\begin{figure}[htb]
\centering
\includegraphics[width=12cm]{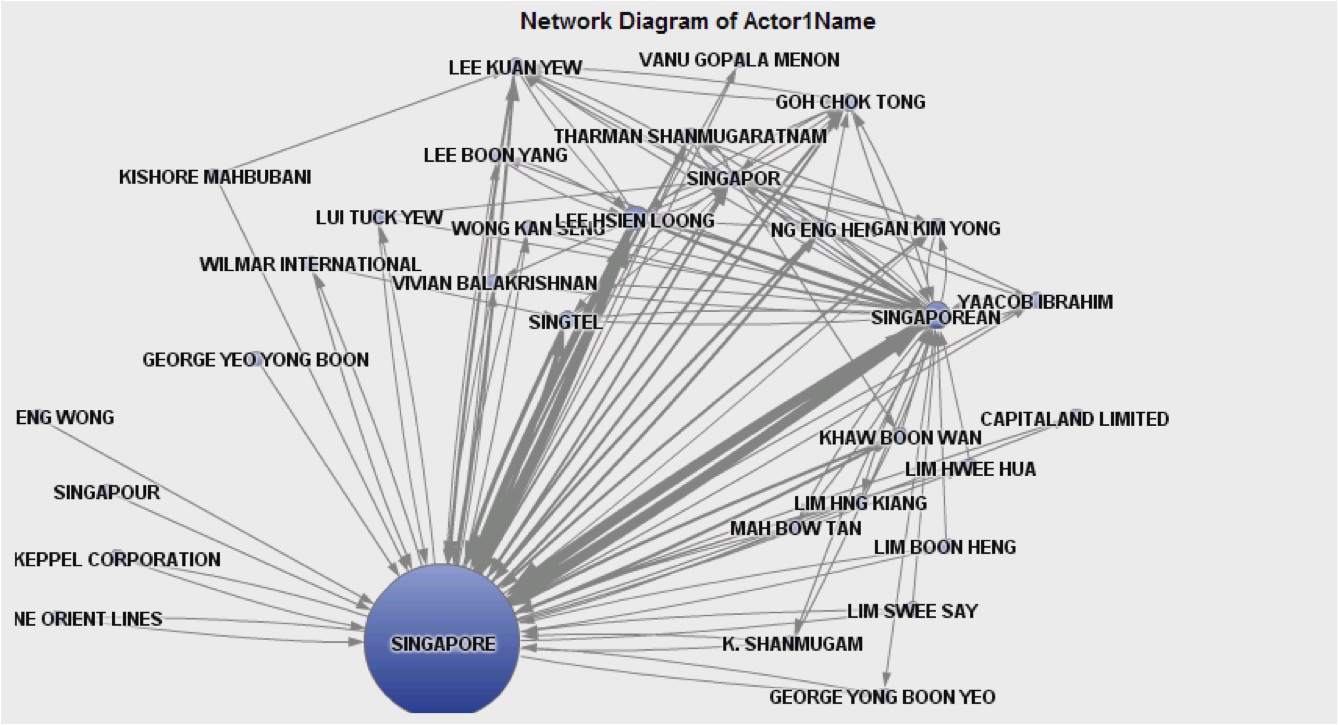}
\caption{Network diagram of names that appear in both \textit{Actor1Name} and \textit{Actor2Name}; and their interactions within Singapore GDELT articles from 2006-2013. Most of the actors are current or past Singapore Cabinet ministers (including K. Shanmugam), with a few companies which are part of the \^{}STI.}
\label{fig:network}
\end{figure}

\begin{figure}
\centering
\includegraphics[width=12cm]{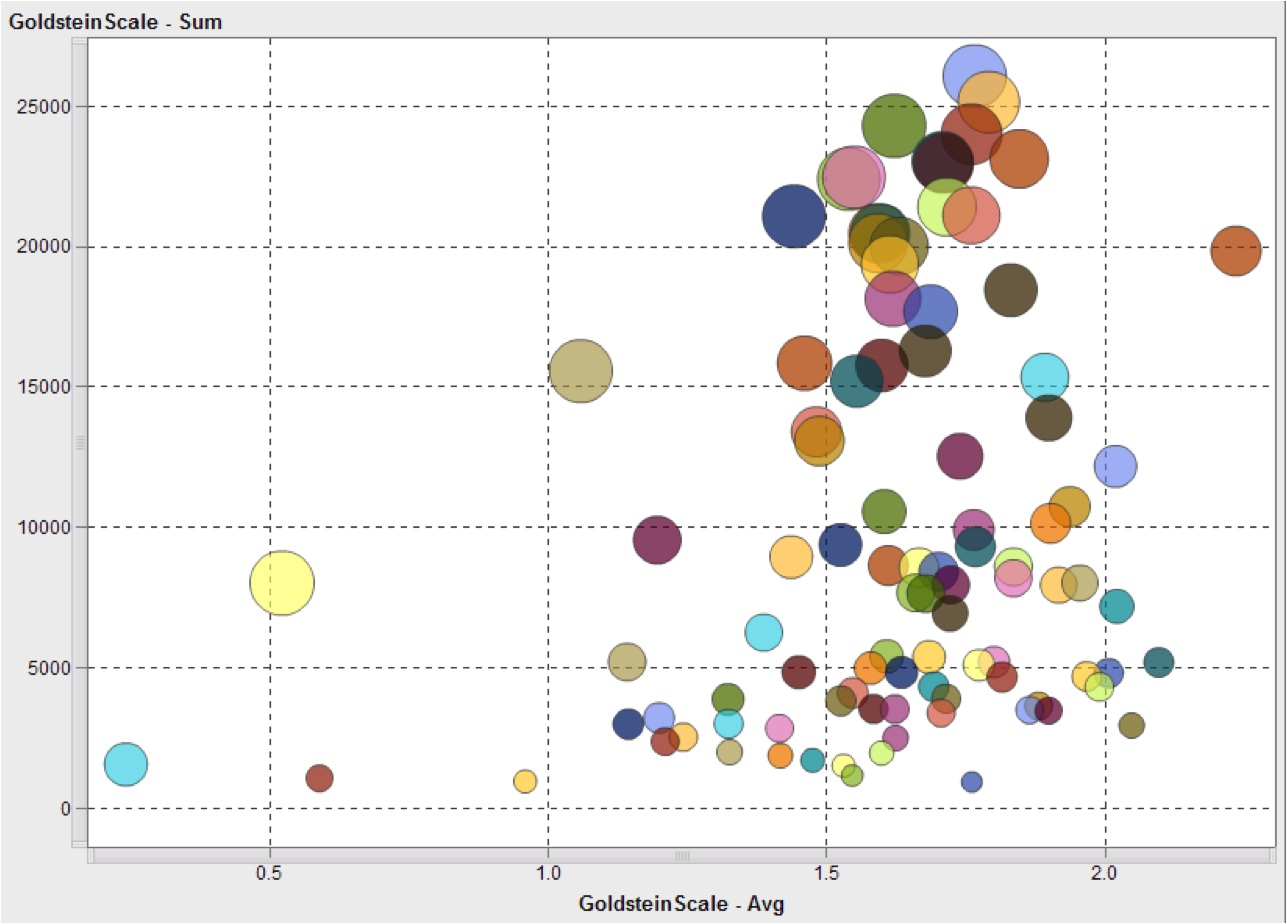}
\caption{Bubble plot on monthly Singapore GDELT articles from 2006-2013. This plot uses all 3 aggregate functions on the \textit{GoldsteinScale}. \textit{x}-axis - \texttt{Avg(GoldsteinScale)}, \textit{y}-axis - \texttt{Sum(GoldsteinScale)}, size of bubble - \texttt{Count(GoldsteinScale)}, and bubble (represents a particular month from 2006-2013). To interpret any such plot, bubbles that are at the extreme top right (most positive impact) or bottom left (most negative impact) would indicate interesting months. This bubble plot shows that Singapore is a relatively peaceful country as the bubbles mostly occupy the positive parts of both axes.}
%-The plot would theoretically in an eventful situation have both positive and negative values at either axis, forming a cross.
\label{fig:gdelt-goldstein-scale-bubble}
\end{figure}

\begin{figure}[htb]
\centering
\begin{minipage}[b]{0.45\linewidth}
\includegraphics[width=5.5cm]{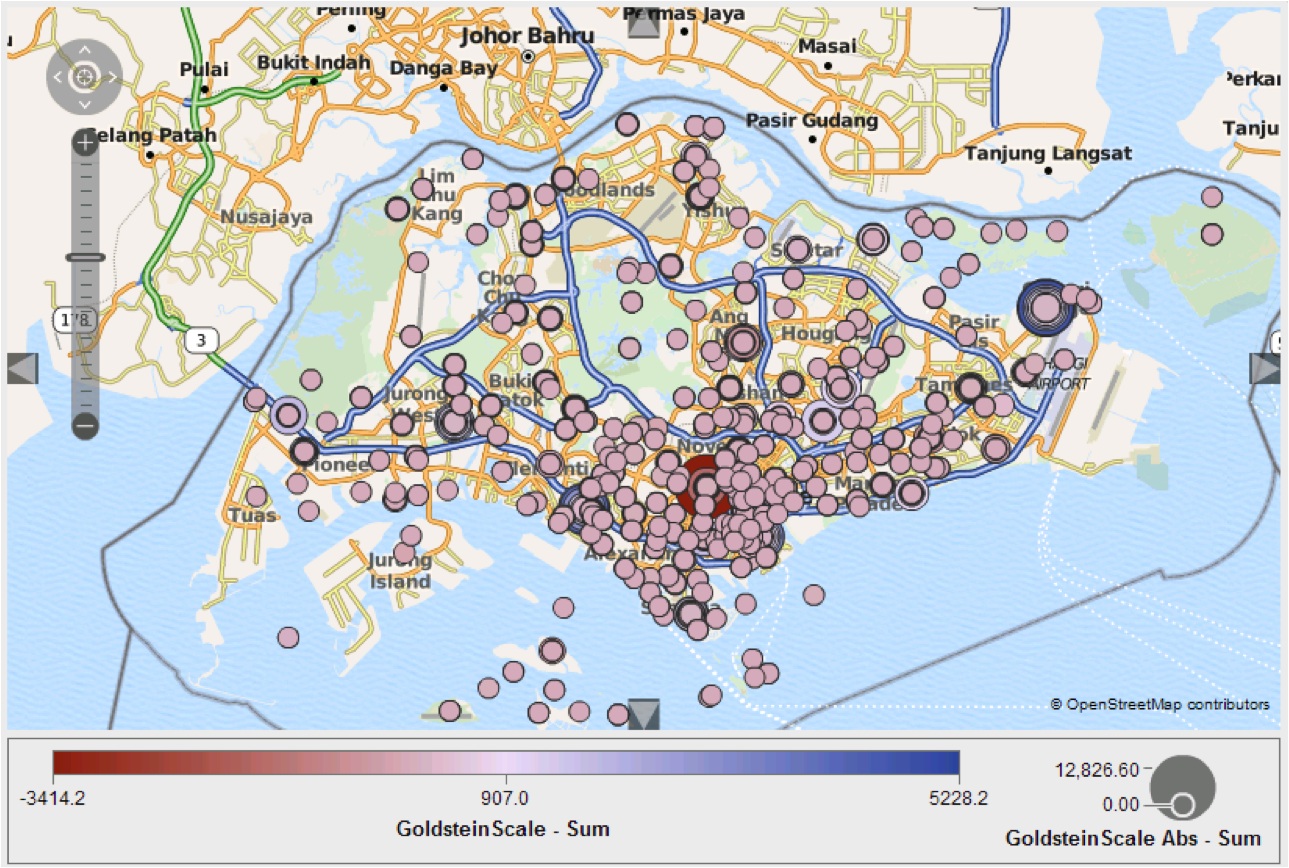}
\caption{Geomap of Singapore GDELT articles from 2006-2013. The size and color of the bubble is dependent on \texttt{Sum(GoldsteinScale)} in that location. More than 90\% of the data points have been filtered out because the locations of those data points was in the center of Singapore (the geographical location captured was simply Singapore and not a specific location within Singapore). Because of the long time period of 8 years, the visualization is too cluttered.}
\label{fig:geomap-2006-2013}
\end{minipage}
\quad
\begin{minipage}[b]{0.45\linewidth}
\includegraphics[width=5.5cm]{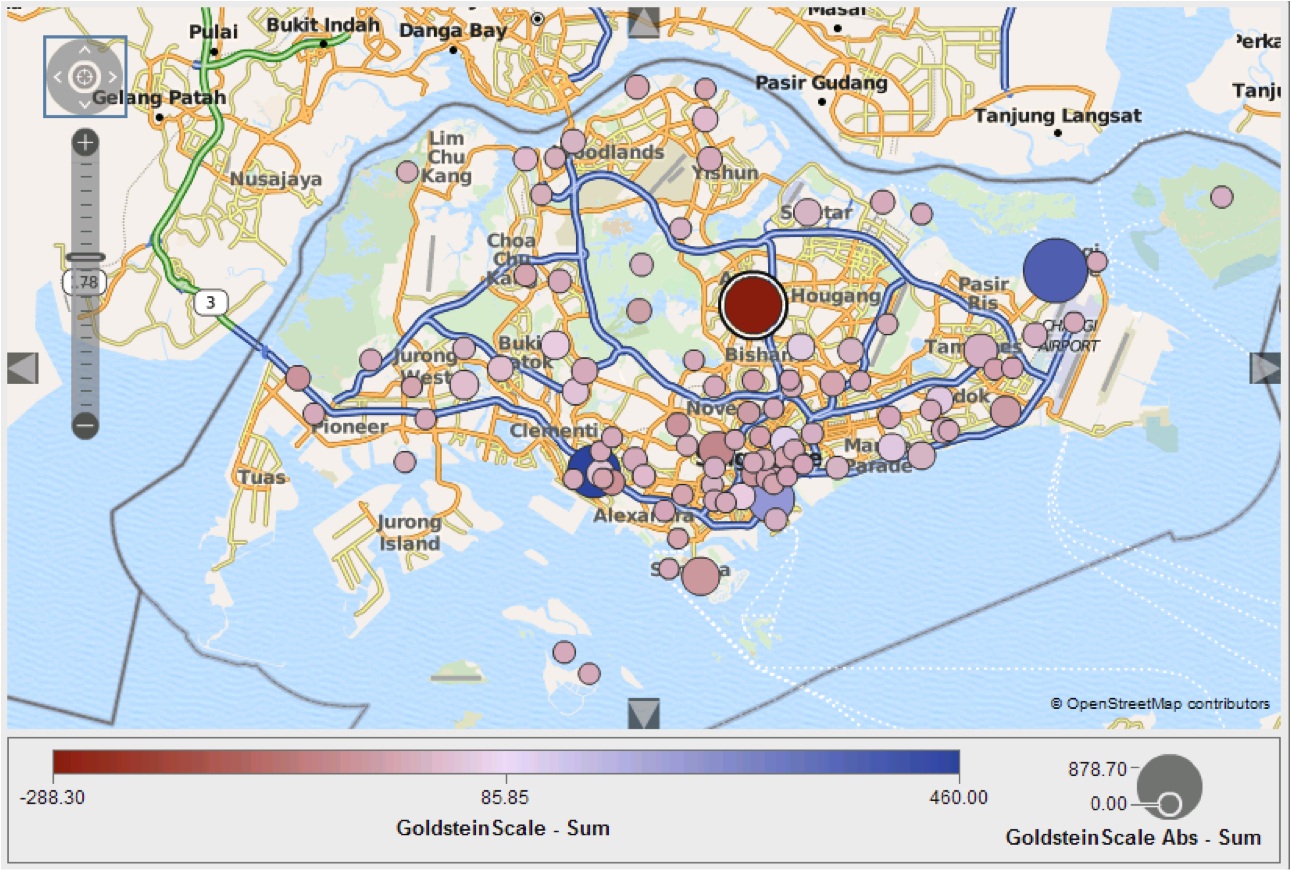}
\caption{Geomap of Singapore GDELT articles from November 2013 (1 month only). With reference to the \textit{sourceURL} column, the highlighted big red bubble is about the Anonymous hacker cyber attack on the \href{http://www.amktc.org.sg/}{Ang Mo Kio town council website} - a key Singapore event; and the big blue bubble near Clementi is due to the opening of a new corporate lab by National University of Singapore (NUS) and Keppel Corporation, and various other NUS scholarship launches - not a key Singapore event.}
\label{fig:geomap-november-2013}
%\begin{figure}
%\centering
%\includegraphics[width=12cm]{Geomap-november-2013-sourceURL.jpg}
%\caption{\textbf{(SHOULDN'T THIS BE A TABLE???)}}
%\label{fig:geomap-november-2013-sourceURL}
%\end{figure}
\end{minipage}
\end{figure}

%\section{Appendix: Accuracy of Term Extraction}

\begin{table}[htb]
\centering
\begin{minipage}[b]{1\linewidth}
\caption{Topic analysis results for 25 topics}
\begin{center}
    \begin{tabular}{|l|l|l|l|}
\hline
Topic                                          & \#Terms & \#Docs & Description \\
\hline
+school,+student,+woman,+child,+man            & 6980    & 10558  &  \\
+company,+business,+technology,+industry,asia  & 5586    & 10389  &  \\
percent,+bank,+market,+company,+billion        & 4607    & 9502   & \\
+minister,+cooperation,lee,prime,asean         & 4578    & 9486   & \\
+ship,+navy,+defense,+region,military          & 5828    & 8309   & \\
+government,political,+policy,+party,+election & 5613    & 8200   & \\
malaysia,datuk,pontian,najib,seri              & 5907    & 8187   & \\
tpp,+trade,+agreement,japan,+negotiation       &         & 7638   & \\
+medium,websites,website,internet,+hack        & 5520    & 7591   & \\
+airline,+carrier,+flight,airlines,+passenger  & 4098    & 7163   & \\
+worker,+riot,+police,india,little             & 4948    & 7016   & Little India riot \\
+court,delhi,+accuse,+victim,+bus              & 3124    & 5183   & \\
+haze,+fire,indonesia,sumatra,pollution        & 4054    & 4651   & Southeast Asia haze \\
todd,shane,+suicide,+coroner,+death            & 3307    & 3903   &  \\
sullivan,clinton,iran,obama,+iranian           & 2239    & 3099   & \\
+oil,palm,+palm,rspo,palm oil                  & 893     & 2484   & \\
francis,+navy,misiewicz,beliveau,+ship         & 2082    & 1856   & \\
tpp,intellectual,wikileaks,+proposal,+property & 574     & 1135   & \\
bernama,jalan,wisma,65a,603-2693               & 358     & 893    & \\
uruzgan,tarin,kot,afghan,afghanistan           & 515     & 787    & \\
hasina,jan 12 2014,sheikh,+swear,+poll         & 1102    & 487    & \\
menafn,khaleej,jan 11 2014,+arab time,arab     & 470     & 478    & \\
cadets,evesham,petty,dofe,littleton            & 347     & 290    & \\
div,+class,li,div class,ul                     & 49      & 263    & \\
camaro,a8,zl1,chevy,+engine                    & 714     & 253    & \\
\hline
    \end{tabular}
\end{center}
\label{table:topic25}
\end{minipage}
\quad
\begin{minipage}[b]{1\linewidth}
\caption{Breaking news identified from the articles}
\begin{center}
    \begin{tabular}{|l|l|}
\hline
Topic Terms                                     & Description \\
\hline
+spy,snowden,intelligence,australia,+allegation & \\
philippines,+typhoon,haiyan,relief,philippine   & \\
+police,+murder,kovan,+suspect,iskandar         & Kovan double murder \\
gay,377a,+man,+dot,+cookie                      &  \\
+riot,+worker,little,+police,india              & Little India riot \\
+haze,+fire,indonesia,pollution,sumatra         & Southeast Asian haze \\
todd,shane,+suicide,+coroner,+death             & \\
tpp,intellectual,wikileaks,+proposal,+property  & \\
\hline
    \end{tabular}
\end{center}
\label{table:breaking}
\end{minipage}
\quad
\begin{minipage}[b]{1\linewidth}
\caption{Number of common topics}
\begin{center}
    \begin{tabular}{|c||c|c|c|c|}
\hline
Total & 100 & 75 & 50 & 25\\
\hline
\#Unique Common Topics & 20 & 18 & 15 & 9\\
\hline
    \end{tabular}
\end{center}
\label{table:common-topic}
\end{minipage}
\end{table}

%\section{Appendix: Potential for Prediction}

\begin{figure}
\centering
\includegraphics[width=5.5cm]{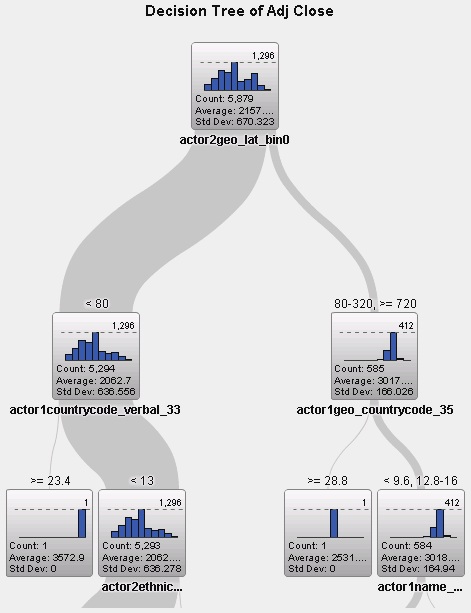}
\caption{Decision tree constructed using advanced growth strategy \cite{sas2013va} (p242), independent variables of Singapore GDELT articles at day $t$, and target variable of \^{}STI adjusted close price at day $t+1$, from 1988-2013 (weekends are excluded - 31\% of data). The top-3 variables (out of 1,100) to predict \^{}STI adjusted close price are \textit{Actor2Geo\_Lat = NULL},  \textit{Actor1CountryCode = Netherlands}, and \textit{Actor2EthnicCode = Scottish}. This result does not seem to much sense to us. All decision tree results reported in this paper are limited because it uses only features created from news article a day before and did not use target variable as input variables (such as for forecasting).}
%Root mean squared error (the smaller the better) is over 500 (EM Decision Tree) and over 3000 (EM Regression) on the validation set \textbf{(TEST SET???)}.
%R-Square        0.2390
\label{fig:overall-decision-tree}
\end{figure}

\end{document}